\documentclass[twocolumn,showpacs,preprintnumbers,amsmath,amssymb]{revtex4}
\usepackage{epsfig}
\usepackage{amsmath}

\begin{document}

\title{Random and Longest Paths: Unnoticed Motifs of Complex Networks}

\author{Luciano da Fontoura Costa}
\affiliation{Institute of Physics at S\~ao Carlos, University of
S\~ao Paulo, PO Box 369, S\~ao Carlos, S\~ao Paulo, 13560-970 Brazil}

\date{17th Nov 2007}

\begin{abstract}
Paths are important structural elements in complex networks because
they are finite (unlike walks), related to effective node coverage
(minimum spanning trees), and can be understood as being dual to star
connectivity.  This article introduces the concept of random path
applies it for the investigation of structural properties of complex
networks and as the means to estimate the longest path.  Random paths
are obtained by selecting one of the network nodes at random and
performing a random self-avoiding walk (here called path-walk) until
its termination.  It is shown that the distribution of random paths
are markedly different for diverse complex network models
(i.e. Erd\H{o}s-R\'enyi, Barab\'asi-Albert, Watts-Strogatz, a
geographical model, as well as two recently introduced path-based
network types), with the BA structures yielding the shortest random
walks, while the longest paths are produced by WS networks.  Random
paths are also explored as the means to estimate the longest paths
(i.e. several random paths are obtained and the longest taken).  The
convergence to the longest path and its properties ire characterized
with respect to several networks models.  Several results are reported
and discussed, including the markedly distinct lengths of the longest
paths obtained for the different network models.
\end{abstract}

\maketitle

\vspace{0.5cm}
\emph{`The longest journey starts with a single step.' 
(Old Chinese proverb.)}

\section{Introduction} 

A good deal of the attention so far focused on complex networks
(e.g.~\cite{Albert_Barab:2002, Dorogov_Mendes:2002, Newman:2003,
Boccaletti:2006, Costa_surv:2007}) has been motivated by properties
involving high degree nodes (hubs in scale free models) and shortest
paths between pairs of nodes (small world networks).  The former model
is characterized by scale free degree distribution, which accounts for
the heterogeneity of the local connectivity of nodes and the presence
of hubs.  Scale free models are also small-world because each hub
establishes shortcuts among all nodes to which it is connected to.
Recently~\cite{Costa_path:2007, Costa_comp:2007}, the concept of a
\emph{star} was suggested in order to represent the network motifs
defined by the edges connected to each node.  As discussed in those
works, a star can be understood as the \emph{dual} of a
path~\footnote{A \emph{walk} is any sequence of adjacent edges along
the network.  A \emph{path} is a special type of walk which never
repeats edges or nodes.}.  Particularly long paths can therefore be
related to hubs through such a duality.  Stars and paths,
inter-related by the path-star transformation and its
inverse~\cite{Costa_path:2007, Costa_comp:2007}, present completely
opposite intrinsic properties: stars are centralized, while paths are
sequential.

While great attention has been given to the degree (e.g. presence of
hubs) and \emph{shortest paths}, other types of paths --- such as
random and longest --- have received little attention.  It is
particularly interesting to study paths because: (i) they have finite
length (unlike walks); (ii) they correspond to the connected subgraph
with the smallest possible number of edges linking a set of nodes (a
concept related to that of minimum spanning tree); and (iii) as
outlined above, they can be understood as duals of stars, a
fundamentally important motif in complex networks.

Random paths, as suggested in this work, can be obtained by selecting
a node at random and performing a random path-walk (i.e. a walk not
allowed to repeat nodes and edges) until unvisited edges or nodes are
no longer available.  Such paths are therefore closely related to the
\emph{exploration}~\cite{Costa_know:2006, Costa_exploring:2007}, 
\emph{sampling}~\cite{Paulino:2007}  and 
\emph{coverage}~\cite{Costa_know:2006, Costa_trails:2007} of the nodes 
in a network while trying to minimize the number of edges crossed (not
a single edge or node in a path are visited more than once).
Actually, it is interesting to observe that maximal paths, such as
Hamiltonian paths~\footnote{Hamiltonian paths are related to the more
widely-known Hamiltonian cycles, in the sense that these two types of
structures can be transformed one into the other by adding/removing a
single edge.} covering all the nodes of a network, are special cases
of minimum spanning trees~\cite{Min_span:2002}. Random paths can also
be understood as being produced by moving agents which destroy the
edges and nodes as they are visited, as it could happen during the
spreading of a disease or attack on a network.  In addition to being
related to such dynamical processes, paths can be considered in order
to get insights about the structure and topological properties of
networks.  For instance, if the several random paths obtained from a
given network are found to have short lengths, it may be inferred that
the network contains many bottlenecks (e.g. hubs or edges with large
betweeness centrality).  Therefore, the analysis of random paths can
provide additional valuable information about the structure of the
analyzed networks.

The longest path in a complex network corresponds to the path (or one
of the equal-length paths) involving the largest number of edges (and
therefore nodes).  Alternatively, the longest path can be defined as
the random path involving the largest number of edges.  While the
distribution of the shortest path lengths bears important implications
for communications and distributions taking place in the network, the
longest path is naturally related to effective coverage of the nodes
by a moving agent.  For instance, the presence of a Hamiltonian path
indicates that all nodes of the network can be visited while crossing
the minimal number of edges.  Therefore, the existence of long paths
in a specific network may indicate that these structures are
intrinsically important for the dynamics or that the network has been
constructed around or by incorporating such paths
(e.g.~\cite{Costa_comp:2007}).  One of the possible reasons why scant
attention has been focused on longest paths is that, while shortest
paths can be determined in polynomial time, longest path
identification is an NP-complete problem (e.g.~\cite{Gibbons:1985}).

The current article aims at investigating random and longest paths in
diverse theoretical complex networks models.  It starts by describing
the basic concepts and methodology, including a simple stochastic
method for estimating the longest path, and proceeds with the
experimental investigation of the random and longest paths in six
theoretical complex networks models.

\section{Basic Concepts}~\label{sec:basic}

A undirected complex undirected network, formed by $N$ nodes and $E$
edges among those nodes, can be fully represented in terms of its
\emph{adjacency matrix} $K$ of dimension $N \times N$.  Each existing
edge $(i,j)$ implies $K(i,j)=K(j,i)=1$, with $K(i,j)=K(j,i)=0$ being
otherwise imposed.  Two edges are said to be \emph{adjacent} whenever
they share one of their extremities.  A \emph{walk} corresponds to any
sequence of adjacent edges $(i_1,i_2); (i_2,i_3); \ldots
(i_{p-1},i_p)$, where $p$ is the \emph{length} of the walk. A walk
which does not repeat any edge or node is a \emph{path}.  An
\emph{Hamiltonian path}~\cite{Diestel:2000} is a path that encompass
all nodes in the networks.  The \emph{length} of a path is equal to
the number of its constituent edges.  The shortest path between two
nodes is defined as one of the paths between those nodes which has the
smallest length.  

The \emph{immediate neighbors} of a node $i$ are those nodes which are
connected to $i$ through shortest paths of length 1.  A \emph{star} is
a motif containing a node $i$ and the edges attached to it.  The
\emph{degree} of a node is equal to the number of edges in its
respective star.  The node degree averaged within a network is called
\emph{average degree}.  The \emph{clustering coefficient} of a node $i$ 
is the ratio between the number of undirected edges between the
immediate neighbors of $i$ and the maximum possible number of
undirected edges among those nodes.

Six theoretical models of complex networks are considered in the
present work including four traditional models --- Erd\H{o}s-R\'enyi
(ER), Barab\'asi-Albert (BA), Watts-Strogatz (WS) and a geographical
model (GG) --- as well as two recently introduced knitted types of
complex networks~\cite{Costa_comp:2007} --- the path-transformed BA
model (PA) and path-regular networks (PN).  The ER, BA and WS networks
are grown as traditionally done (e.g.~\cite{Albert_Barab:2002,
Dorogov_Mendes:2002, Newman:2003, Boccaletti:2006, Costa_surv:2007}).
The GG network is obtained by randomly distributing $N$ nodes within a
square and connecting those nodes which are closer than a parameter
$d$.  The PA and PN networks are grown as described
in~\cite{Costa_comp:2007}.  Basically, the PA networks
(\emph{path-transformed BA} networks) are obtained by star-path
transforming all nodes in an original BA network.  The PN (\emph{path
regular} networks) model, on the other hand, is easily obtained by
defining paths involving all network nodes in random order and without
repetition.

All networks have $N=100$ and $m=3$ or $m=5$ ($m$ is the number of
spokes in the added nodes in the BA model), which implies average
degree $\left< k \right> = 2m$.  Because the average degrees
considered in this work are above the percolation critical density,
most of the nodes in each network belong to the largest connected
component, which is the only part of the network considered for the
analyses reported in this article.

\section{Methodology}~\label{sec:method}

This section describes the simple stochastic algorithm for estimation
of the longest path used this article as well as the several
path-based measurements which can be applied in order to characterize
the global connectivity of networks.  Given a network, the longest
path estimation approach involves finding many random paths while
keeping track of the longest obtained path length.  Despite the fact
that the determination of the longest path is NP-complete, the method
is verified, at least for the considered networks, to converge steady
and quickly to the longest path.

Given a complex network, one may choose a node $i$ at random and start
a self-avoiding random walk (or path-walk~\footnote{Because a path is
usually applied to a static structure in a network, here we use the
name \emph{path-walk} in order to express the process of obtaining a
path through a self-avoiding random walk.  Observe that the result of
a path-walk consequently is a path.}) from two outbound edges of $i$,
until unvisited edges and nodes are no longer available to the walker
at both sides of the walk (a situation which will be called here
\emph{saturation or termination of the path}).  It is proposed in this 
work that such a \emph{random path} can provide valuable information
for complex networks studies.  More specifically, provided several
random paths are obtained by using such a simple algorithm, the
distribution of their properties, especially their
\emph{length}, can provide valuable information about the global 
structure of the network connectivity.  

The algorithm adopted for picking a random path is simple and involves
the following steps:

\begin{trivlist}
\item {\bf (i)} Select a node $i$ at random; 
\item {\bf (ii)} Chose two edges $A$ and $B$ emanating from $i$ (in case 
that node has degree one, chose only one edge);
\item {\bf (iii)} Perform a path-walk along both edges $A$
and $B$ until the walk stops (i.e. a node is reached which leads to no
unvisited node or edge).
\end{trivlist}

The length of a random path is henceforth represented by $p$.

The algorithm proposed here for the estimation of the longest path is
also very simple and involves obtaining several random paths and
taking the one exhibiting the largest length as the \emph{longest
path}.  Given that the determination of the longest path is
NP-complete, it is important to consider the convergence of such a
simple algorithm.  This is performed experimentally in
Section~\ref{sec:conv}.

\section{Results and Discussion}

The experimental investigations reported in this section, reported in
the following subsections, involved 50 realizations of each model (PA,
PN, ER, BA, WE and GG) with $N=100$ nodes and considering and $m=3$
and $m=5$.

\subsection{Random Paths in Network Models}

\begin{figure*}
  \vspace{0.3cm}
  \centerline{\includegraphics[width=1.0\linewidth]{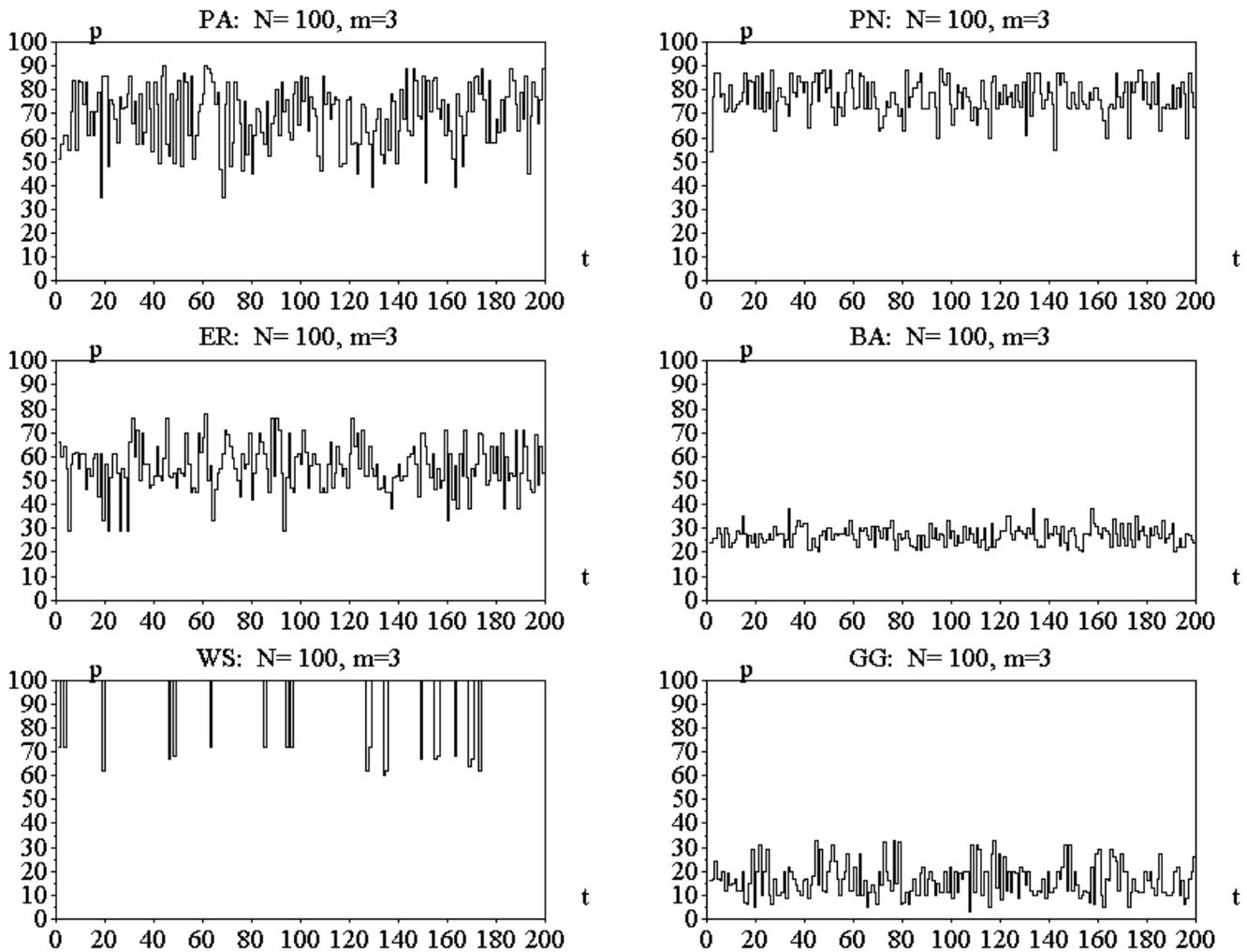}}
  \caption{Lengths of the first 200 randomly selected 
      paths obtained for each of the considered networks models.}
      \label{fig:lengs_time}
\end{figure*}

The first important point to be investigated regards the properties of
random paths extracted from the several theoretical models of complex
networks considered here, (i.e. PA, PN, ER, BA, WS and GG).  In order
to understand random paths in these several models, random paths were
extracted from a network ($N=100$ and $m=3$) of each type by using the
algorithm described in Section~\ref{sec:method} and their lengths
$p$ determined.  Figure~\ref{fig:lengs_time} shows the length of the
first 200 obtained random paths.  Although this figure considers just
one randomly chosen network of each type, it was verified that similar
results are obtained for other network samples, except for the WS
network, which tended to produce diverse signatures for each specific
network.  It is clear from this figure that the lengths of the random
paths varied substantially for each type of network. The first
interesting result is that the values of $p$ present relatively small
variation, except mainly for the PA and ER models; the BA, PN and GG
structures tended to produce random paths with similar lengths. The GG
and BA networks led to the shortest random paths, while the ER and PA
were characterized by medium size lengths.  The longest lengths were
obtained for the PN, PA and WS models.  Observe also that the standard
deviation of the random path lengths tend to increase with the
respective average values.

\begin{figure*}
  \vspace{0.3cm}
  \begin{center}
  \includegraphics[width=0.45\linewidth]{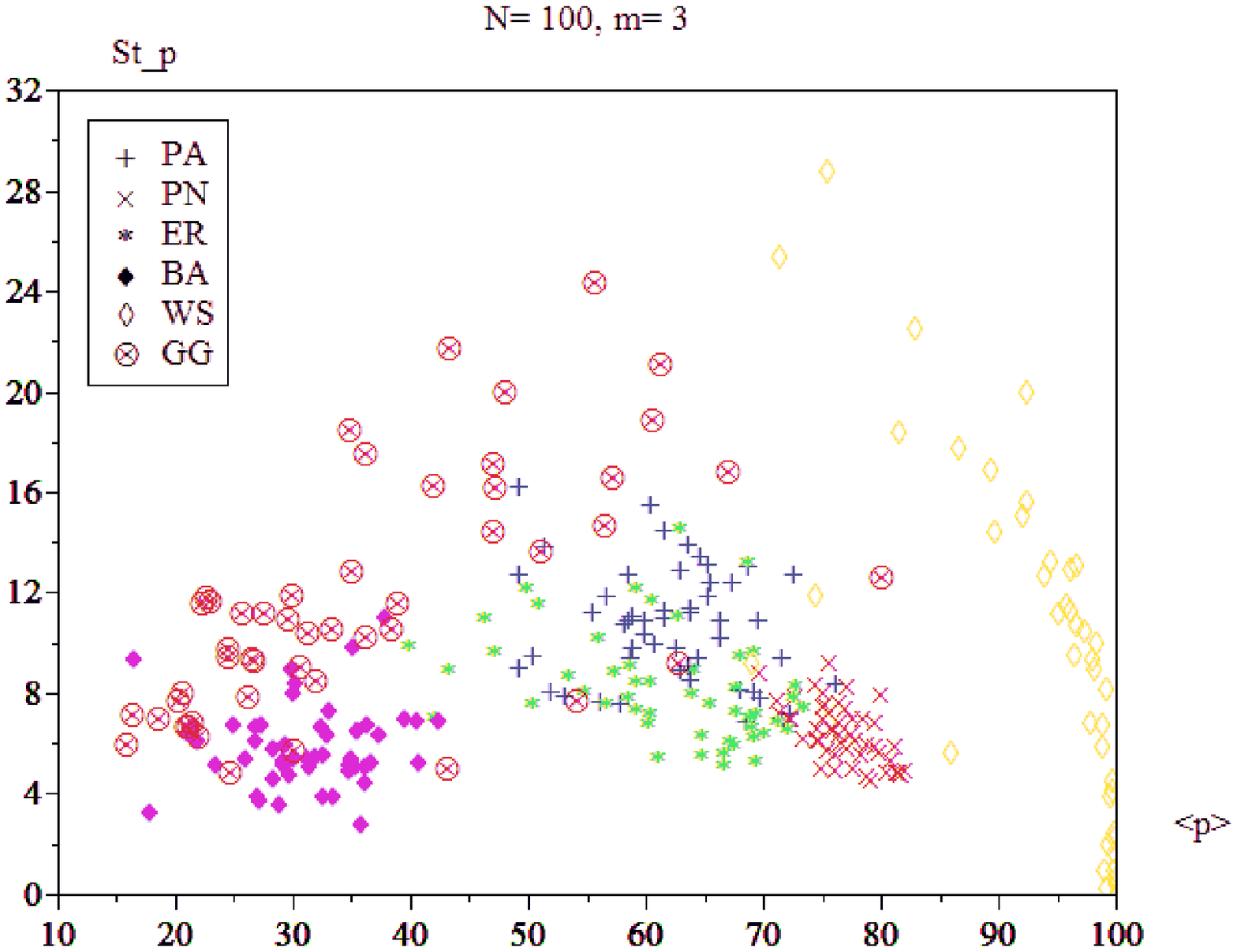}
  \includegraphics[width=0.45\linewidth]{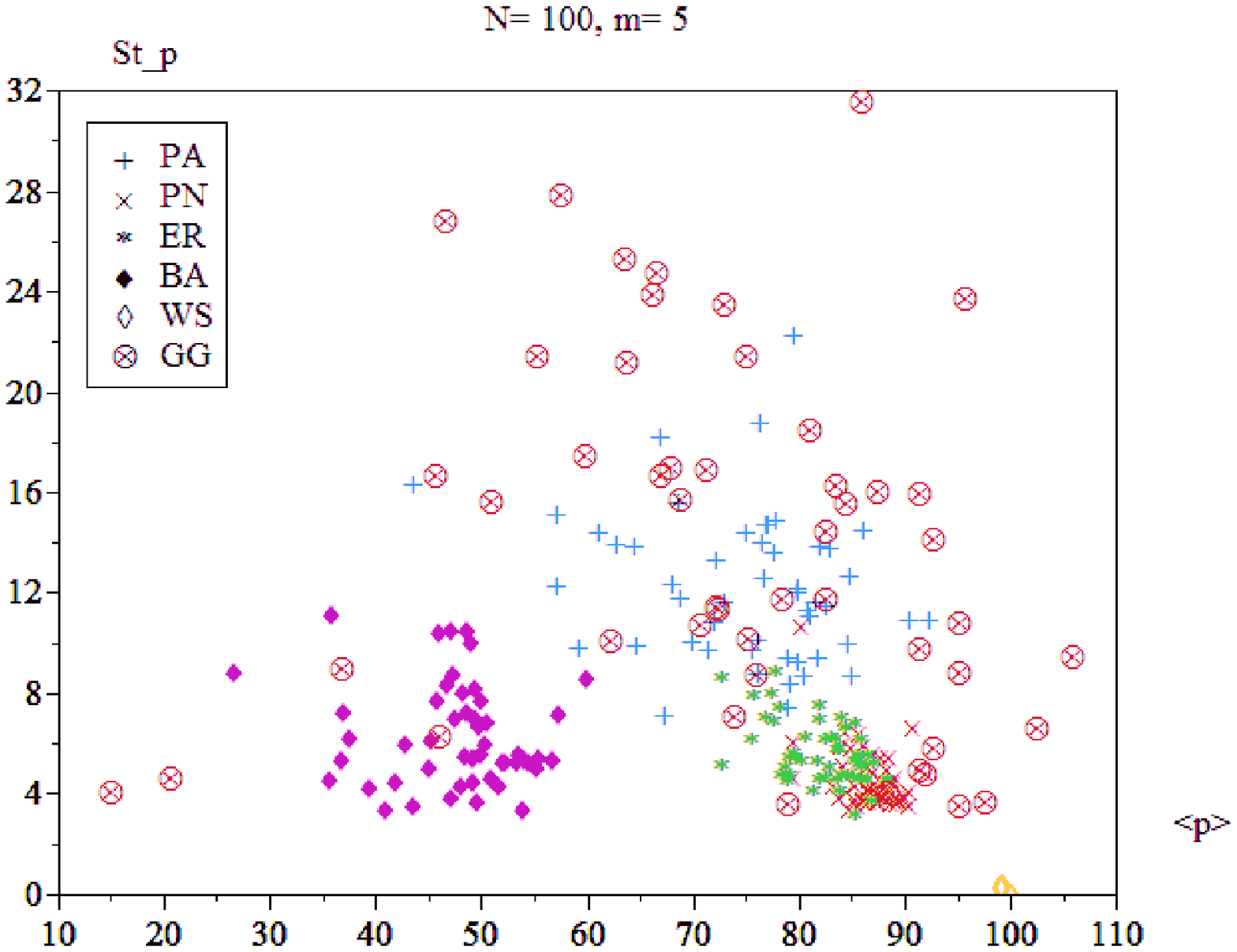}
  \caption{The average and standard deviations of the lengths of 2500
           random paths selected for 50 realizations of each model considering
           $N=100$ and $m=3$.}~\label{fig:scatts_p}
  \end{center}
\end{figure*}

Before interpreting the variations of the random path lengths for each
model, an additional, more systematic study was performed considering
the mean and standard deviation of the length $p$ of 2500 random paths
for $N=100$ and $m=3$ and $m=5$. Figure~\ref{fig:scatts_p} shows the
respectively obtained scatterplots.  Observe that each point in these
scatterplots was obtained from the measurement of the mean and
standard deviation of $p$ considering 2500 random paths from each of
the 50 realizations performed for each configuration (i.e. $N=100$ and
$m=3, 5$).

A series of interesting effects can be inferred from these
scatterplots.  Let us discuss first the case $m=3$, shown in
Figure~\ref{fig:scatts_p}(a).  The BA networks, shown in filled
diamonds, yielded the smallest average value of $p$, followed by the
GG, ER, PA, PN and WS models.  The BA structures are also
characterized by relatively small standard deviations of $p$.  In
other words, paths randomly extracted from BA networks tend to be
short (about 30 edges long) and with similar lengths.  On the other
hand, random paths in WS networks tend to be long (from 70 to 100
edges) and with widely varying lengths.  The PA and ER networks
presented remarkably similar measurements, indicating that these two
models are very similar as far as random path properties are concerned
for $m=3$.  The PN structures resulted remarkably uniform, in the
sense that the mean and standard deviation of $p$ resulted
particularly similar for each of the 50 realizations, defining a
compact cluster in the scatterplot in Figure~\ref{fig:scatts_p}.  The
GG networks yielded widely varying random path lengths extending from
10 to 80 edges.

We now proceed to the scatterplot obtained for $m=5$, i.e. $\left< k
\right> = 10$.  Such an increase of average degree implied some changes 
relatively to the situation discussed for $m=3$.  First, the random
path lengths increased for all network models.  The BA cluster, for
instance, shifted from values in the range $[15, 45]$ to $[25, 65]$.
The largest increase of random path lengths was achieved for the GG
model, which now extends to the maximum length of 100 and over.
Observe that, because of the way in which they are grown, GG networks
may sometimes exceed 100 nodes (see Section~\ref{sec:basic}).
Another interesting effect caused by the average degree increase
regards the separation of the PA and ER cases, with the respective
clusters now presenting substantially less overlap.  Also, the
dispersions of the PN and WS networks were substantially reduced.

Let us know try to discuss each of the main findings above, which is
done as follows:

\vspace{0.2cm}
{\bf (a)}\emph{Longest random paths are obtained for larger average
degree:} Usually, the higher the average number of edges in a network
(reflected in a high average node degree), the more possibilities the
agent performing the path-walk will have to keep proceeding to
unvisited nodes and edges.  Observe that the extreme case where the
network is fully connected (i.e. each node is linked to all other
nodes in the network), all random paths will always have maximum
length of $N$.

\vspace{0.2cm}
{\bf (b)}\emph{Short random paths in the BA model:} Consider a typical
BA network, with its respective hubs.  Once started, a path-walk is
very likely to pass through hubs.  Once such hubs are visited, they
can no longer be part of a path.  Because several of the network edges
are connected to hubs, the paths soon become saturated.

\vspace{0.2cm}
{\bf (c)}\emph{Long random paths in the WS model:} Recall that the WS
model is grown by starting with a regular network (a ring) where each
node is connected to a number of previous and subsequent nodes.
Therefore, by distributing the connectivity almost regularly amongst
the nodes, the chances of having at least one unvisited edge leading
to an unvisited node increases substantially.  Interestingly, being
small world does not imply short random paths.  On the other hand,
the large variances of $p$ obtained for this network model are
accounted by the rewirings used to induce the small-world effect.
Because several of the paths in this model encompass all nodes, this
type of network can be understood to frequently contain Hamiltonian
paths, implying high efficiency of node coverage by moving agents.

\vspace{0.2cm}
{\bf (d)}\emph{Uniformity of random path lengths in the PN model:} As
shown in~\cite{Costa_comp:2007}, PN networks are highly regular
regarding several topological measurements.  Such an enhanced
topological regularity tends to produce similar random path lengths.
Still, the substantially distinct values of average and standard
deviation of the random path lengths obtained for the WS and PN
suggest some intrinsic property for each of those models as far as
random paths are concerned.

\vspace{0.2cm}
{\bf (e)}\emph{Similarity of random path lengths in PA and ER models:}
This effect is particularly intense for $m=3$ (see
Figure~\ref{fig:scatts_p}a).  Interestingly, these two models had been
found~\cite{Costa_comp:2007} to present similar values of several
topological properties.  Such an intrinsic structural congruence
between the BA and PA models seems to have implied also in similar
random path lengths.  Given that the PA model is obtained by applying
the star-path transformation over a BA model, it is surprising that
such a type of networks would result similar to the ER model in so
many aspects.

\vspace{0.2cm}
{\bf (f)}\emph{Large dispersion of random path lengths in the GG
model:} This effect is particularly interesting given that the way in
which such networks are construct could be expected to result in
nearly degree regular networks (i.e. most nodes with similar degrees
because of the randomly uniform spatial distribution of nodes).
However, the random paths obtained from the GG model exhibited widely
varying lengths extending from too short (about 5 edges) to very long
(over 80 edges).  No other considered model yielded such a large
variance of random path lengths.  This property is interesting because
it implies that it is in principle very difficult to foresee how well
a geographical model (at least of the GG type) can be effectively
covered by a moving agent.  In order to understand better the reason
for such a large variance of path lengths in this model, it would be
interesting to consider configurations involving larger values of $d$
(see Section~\ref{sec:basic}).
\vspace{0.2cm}

\subsection{Convergence to the Maximum Random Path}~\label{sec:conv}

Having investigated random paths in the considered complex networks
models, we now proceed to study the longest paths in those structures.

\begin{figure*}
  \vspace{0.3cm}
  \begin{center}
  \includegraphics[width=0.9\linewidth]{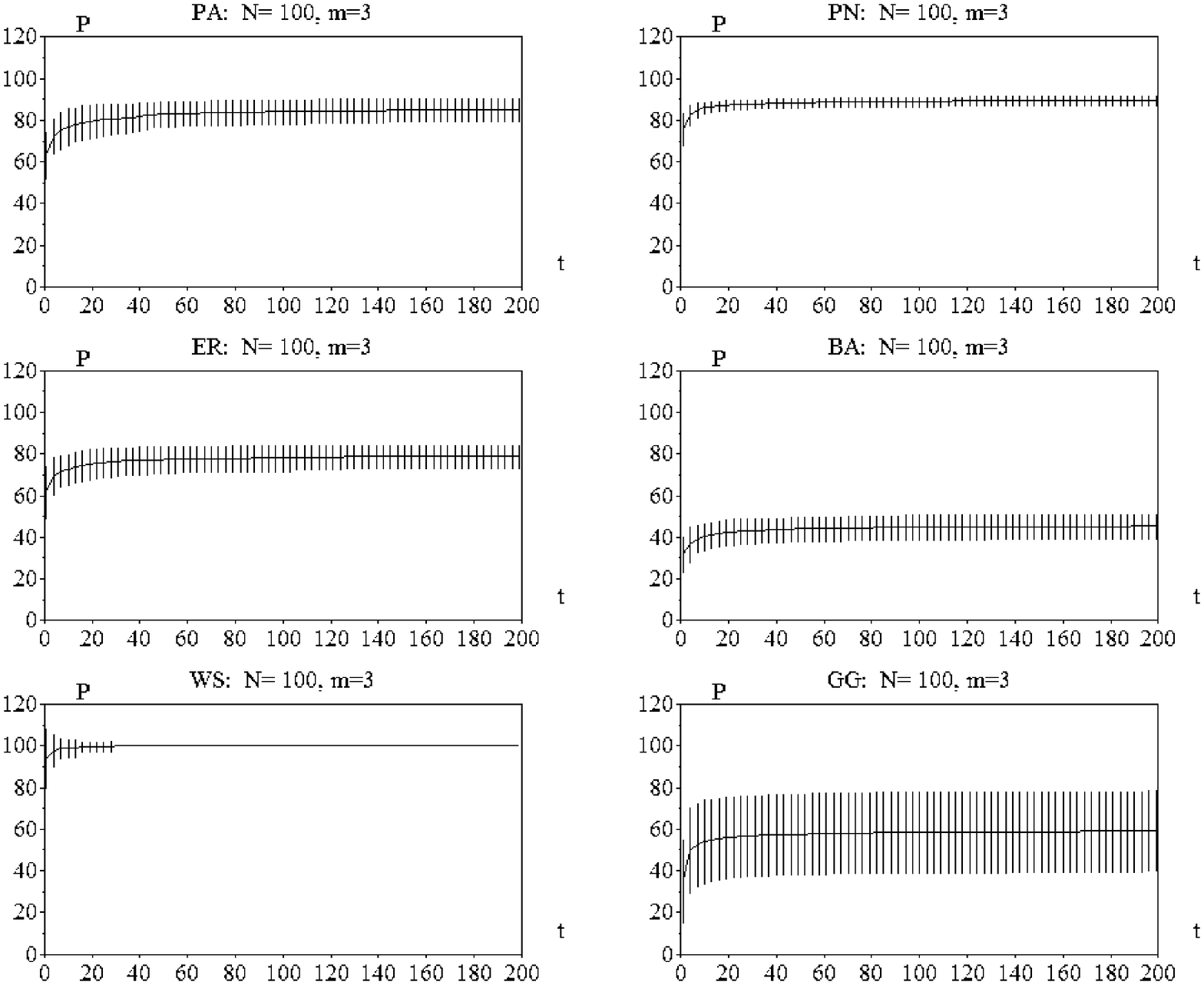}
  \caption{The average $\pm$ standard deviation of the lengths of the
  longest path obtained for 50 realizations of each of the considered
  models with $N=100$ and $m=3$.  Observe the steady convergence to
  the plateau associated to the longest path.}~\label{fig:P_time_3}
  \end{center}
\end{figure*}

Figure~\ref{fig:P_time_3} illustrates the length of the longest random
path obtained along the following of successive random paths in a
single network of each considered type with $N=100$ and $m=3$
considering the initial 200 random paths.  It is clear from these
examples that the length of the currently maximum random path tends to
increase fast and steadily to a stable value, defining a plateau,
which tends to correspond to the maximum random path length $P$.
Experiments involving 2500 steps confirmed that the plateaux tend to
remain stable, suggesting that the longest path has indeed been
identified after 200 steps.  Though the standard deviations vary from
one case to another, they are mostly small (except for the GG
networks).

\begin{figure*}
  \vspace{0.3cm}
  \begin{center}
  \includegraphics[width=0.9\linewidth]{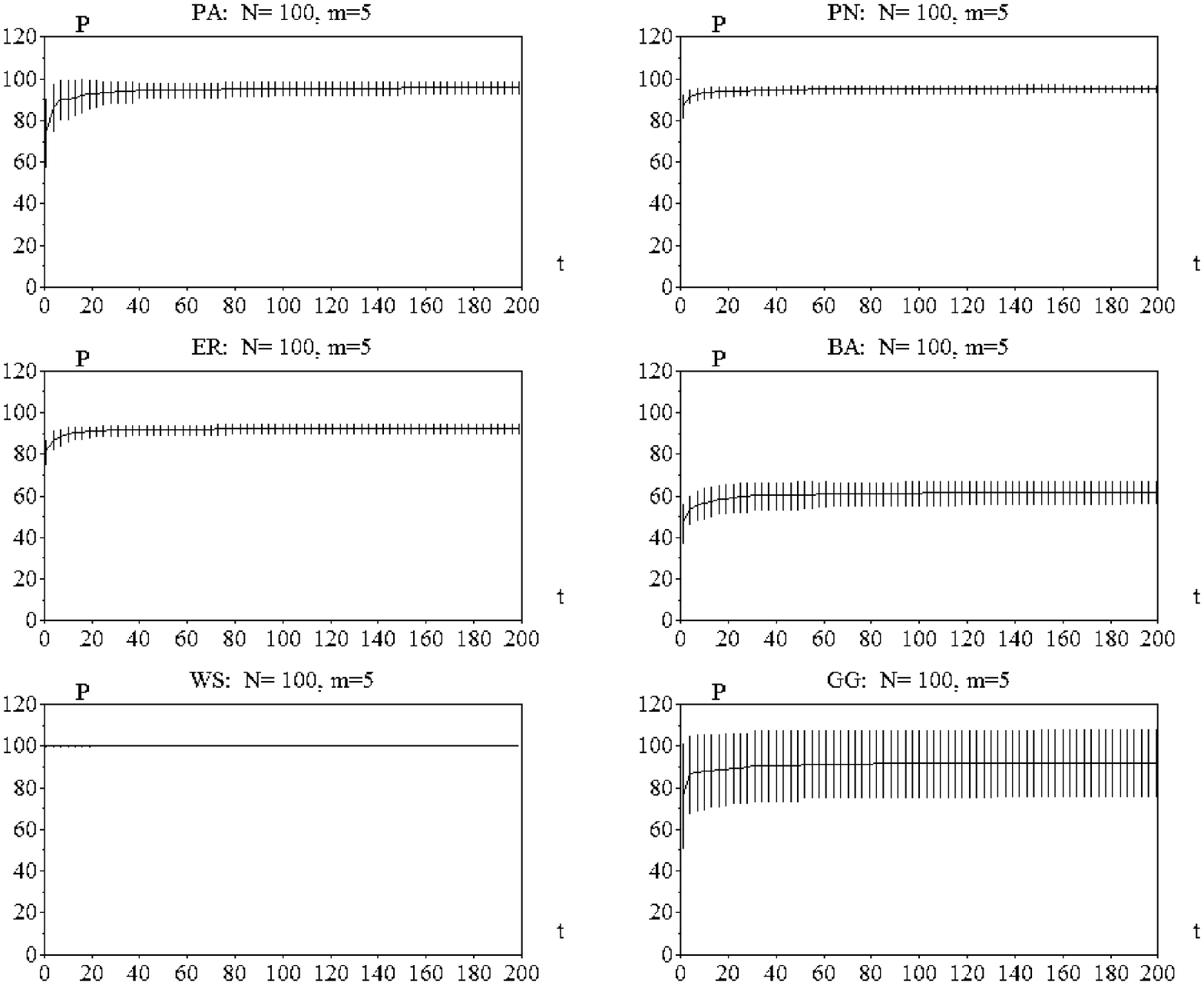}
  \caption{The average $\pm$ standard deviation of the lengths of the
  longest path obtained for 50 realizations of each of the considered
  models with $N=100$ and $m=5$.  Observe the effect of the increase 
  in the average degree ($\left< k \right> = 2m$) in the extension
  of the longest path in all cases, particularly for the GG model.}
  ~\label{fig:P_time_5}
  \end{center}
\end{figure*}

Figure~\ref{fig:P_time_5} presents the longest path lengths for
$N=100$ and $m=5$ along the 200 steps (recall that each step involving
obtaining a random path and comparing its length with that of the
longest current path).  Observe that the increase in $m$, implying
higher average degree, tended to augment the length of the longest
path in all cases, particularly for the GG structures where the
longest path length almost doubled.  It is also interesting to notice
that the standard deviations tended to decrease in all models for this
higher value of $m$.

The above results suggest that the simple methodology adopted in this
work for identification of the longest path tends to converge fast
to the longest path (or at least to a path very similar to that in
length) at least for the considered values of $N$ and $m$.  Also, the
obtained results indicate that most models, except mainly for the BA
networks, contain longest paths passing through the great majority of
the network nodes.  Several Hamiltonian paths have been found for WS
networks.

\subsection{Similarity between random paths and the maximum random path}

Though the method for longest path identification tends to converge
fast and steadily to a putative longest path, possibly representing
the longest one in the network, it is interesting to consider how much
each of these putative longest paths followed during the algorithm
resembles the longest path obtained for a very large number of steps
(henceforth called the \emph{reference longest path}).  The importance
of such an investigation becomes clear when we consider that a single
network may have more than one longest path (with the same or very
similar lengths), possibly even involving the same nodes.  Provided we
can quantify the difference between the reference longest path and
each considered random path, the study of the variation of such a
difference along the steps can supply a valuable information about the
eventual multiplicity of longest paths in the network.  As an example
of a extreme situation, a completely connected network will contain a
large number of Hamiltonian paths, i.e. paths covering all nodes in
the network.

In this section we report experiments aimed at quantifying the
difference between the subsequent random paths and the reference
longest path considering the overlap of nodes and edges.

Given two paths $P1$ and $P2$, each with $N1$ and $N2$ nodes (and
therefore $E1=N1-1$ and $E2=N2-1$ edges), it is possible to estimate
the dissimilarity between these two paths through the following index

\begin{equation} \label{eq:Sn}
  D_n  = \frac{M}{max(N1,N2)}
\end{equation}

where $M$ is the number of nodes which are different in both paths.
Therefore, the index $D_n$ varies between 0 and 1, achieving its
minimum value for identical paths.  However, such a minimum value does
not imply that the two paths are identical (though they have the same
nodes), because each of them may involve different edges from the
original network.

It is also possible to define a dissimilarity index considering the
overlap of edges along the two paths, i.e.

\begin{equation}  \label{eq:Se}
  D_e  = \frac{Q}{max(E1,E2)}
\end{equation}

where $Q$ is the number of different edges in both paths.  Again, we
have that $0 \leq S_e \leq 1$.  The minimum dissimilarity between the
edges of the two paths again provides no guarantee that the two paths
are identical because the order of the edges is not taken into account
in Equation~\ref{eq:Se}.  Though it would be possible to obtain a
completely strict measurement of similarity between the two paths by
considering distances between the two respective adjacency matrices
representing the paths, such a measurement would not work for cases
where the two paths involve the same nodes but different edges.  For
such reasons, the similarities between each random path and the
reference longest path are henceforth performed in terms of the
dissimilarity indices given in Equations~\ref{eq:Sn} and ~\ref{eq:Se}.

\begin{figure*}[h]
  \vspace{0.3cm}
  \begin{center}
  \includegraphics[width=0.9\linewidth]{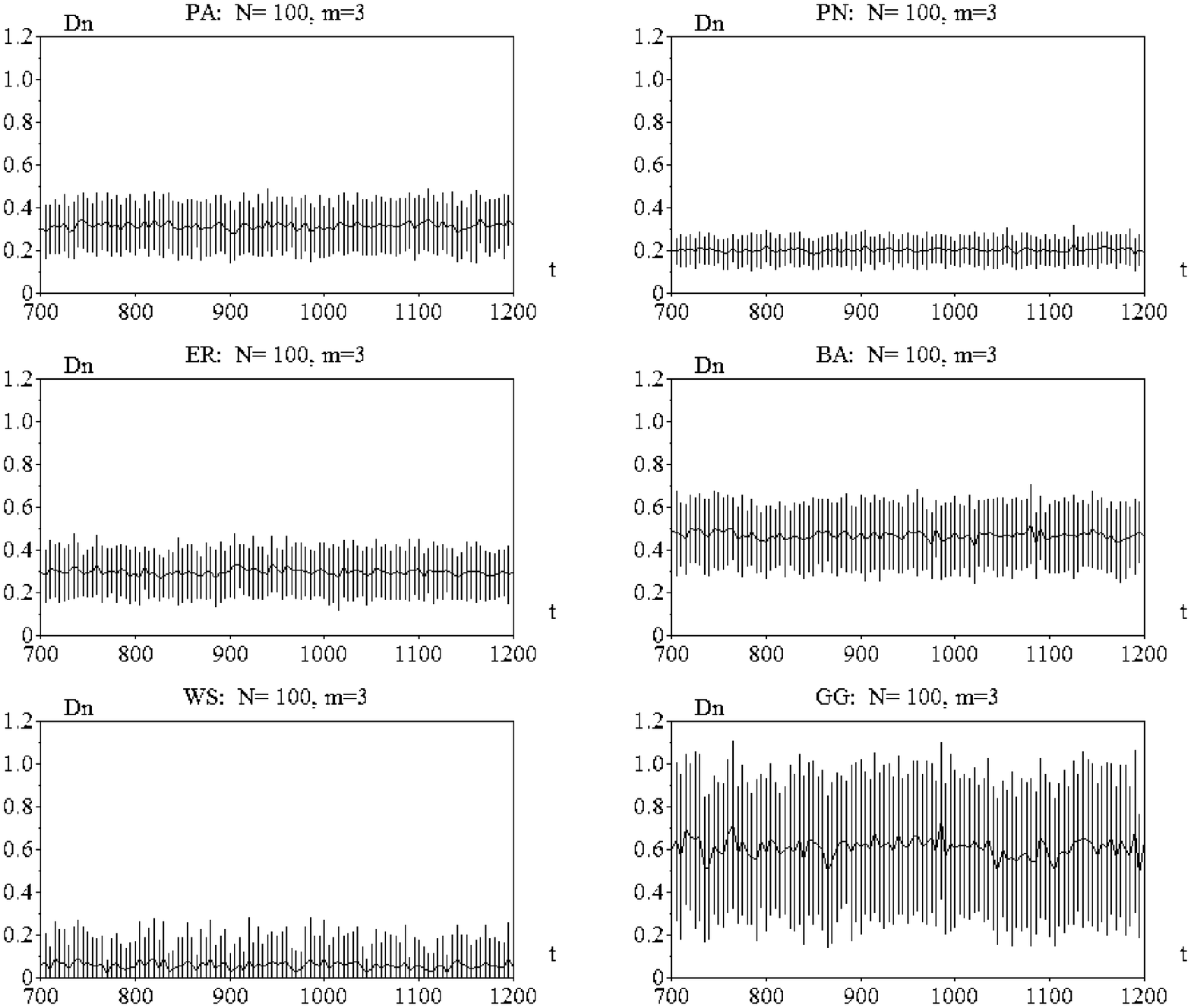}
  \caption{The node dissimilarity between the subsequent
  random paths at each step and the reference path for
  $N=100$ and $m=3$.}~\label{fig:Dn_3}
  \end{center}
\end{figure*}

\begin{figure*}[h]
  \vspace{0.3cm}
  \begin{center}
  \includegraphics[width=0.9\linewidth]{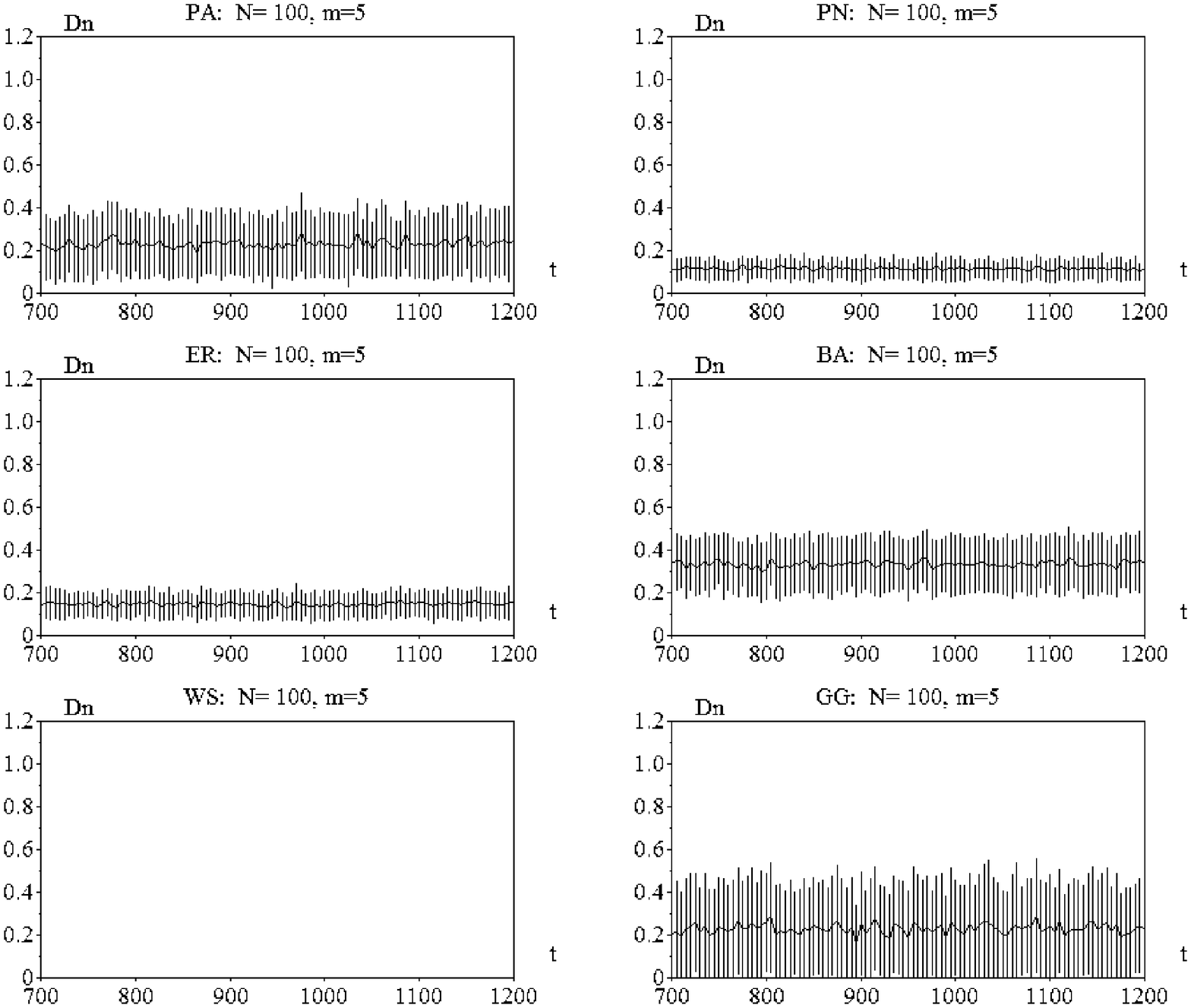}
  \caption{The node dissimilarity between the subsequent
  tandom paths at each step and the reference path
  for $N=100$ and $m=5$.}~\label{fig:Dn_5}
  \end{center}
\end{figure*}

Figure~\ref{fig:Dn_3} shows the node dissimilarity for all cases,
considering $N=100$ and $m=3$, along 500 steps (from step 700 to 1200)
of the algorithm for longest path estimation.  The reference longest
path corresponds to that obtained after 2500 steps in each case.  The
results in this figure indicate that each network model is
characterized by different dismilarities.  The smallest dissimilarities
(about $7\%$) were obtained for the WS networks.  Combined with the
fact that the WS tends to lead to long random paths, the above result
means that the longest paths in these structures tend to incorporate
the same nodes.  The largest dissimilarities (both average and standard
deviations) were obtained for the geographical model, reflecting the
large dispersion of random path lengths already observed for that type
of networks.  Figure~\ref{fig:Dn_5} shows the node dissimilarities
obtained for $m=5$, all of which smaller than the respective
counterparts in Figure~\ref{fig:Dn_3}.

\begin{figure*}[h]
  \vspace{0.3cm}
  \begin{center}
  \includegraphics[width=0.9\linewidth]{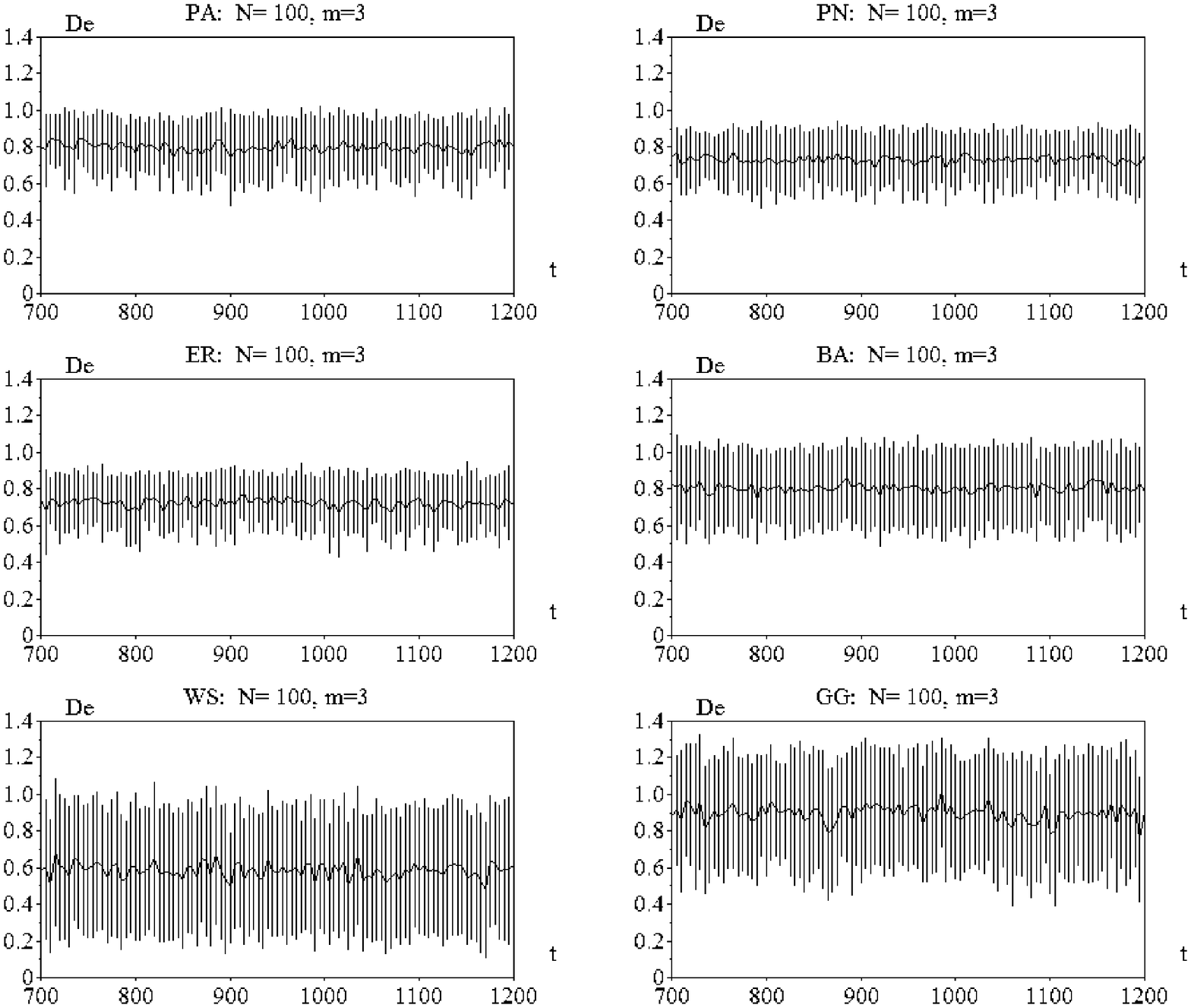}
  \caption{The edge dissimilarity between the subsequent
  random paths at each step and the reference path for
  $N=100$ and $m=5$.}~\label{fig:De_3}
  \end{center}
\end{figure*}

\begin{figure*}[h]
  \vspace{0.3cm}
  \begin{center}
  \includegraphics[width=0.9\linewidth]{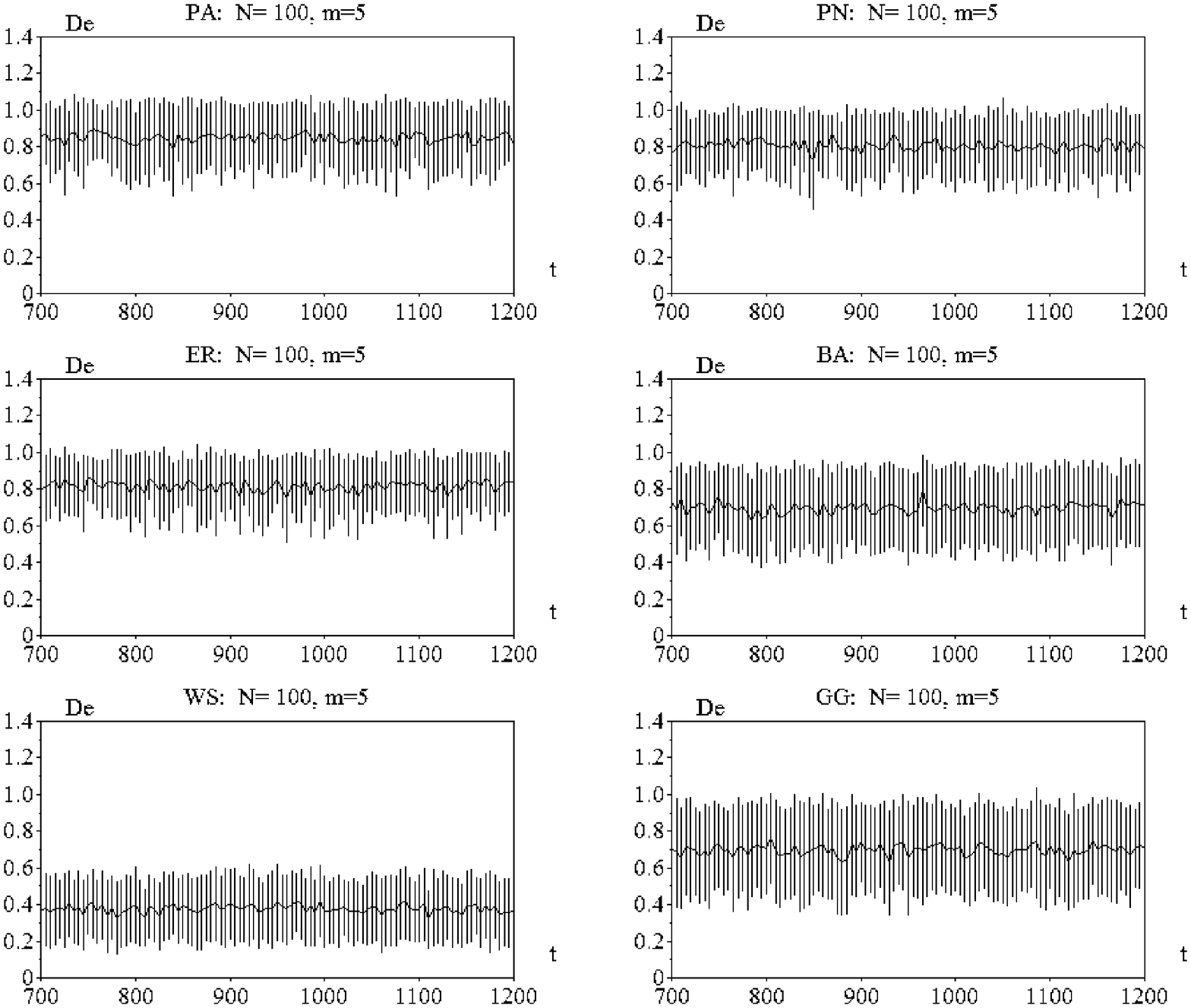}
  \caption{The edge dissimilarity between the subsequent
  random paths at each step and the reference path
  for $N=100$ and $m=5$.}~\label{fig:De_5}
  \end{center}
\end{figure*}

The edge dissimilarities for $n=100$ and $m=3$ are shown in
Figure~\ref{fig:De_3}.  All such relative values are higher than those
respectively obtained for the node dismilarity (please refer to
figure~\ref{fig:Dn_3}).  It is also clear from Figure~\ref{fig:De_3}
that, except for the WS model, the edges dissimilarities were all close
to $80\%$. Therefore, despite the fact that the random paths tend to
share several nodes with the longest path in those networks, they
involve significantly different edges.  The edges similarities
obtained for $n=100$ and $m=5$ are presented in Figure~\ref{fig:De_5}.
These results are remarkably similar to those in
Figure~\ref{fig:De_3}, except for the lower dissimilarity obtained for
the WS model.  Such a result suggests that the average degree seems to
affect relatively little the edge dissimilarities obtained for the
considered models, at least for $N=100$.

\section{Concluding Remarks}

The node degree and shortest path length have received special
attention from the complex network community as measurements of the
topology of complex networks because of their intimate relationship
with the critical properties of scale-free and small-world,
respectively.  Though important, such measurements are not sufficient
to describe the complete topology of complex networks, in the sense
that many distinct networks may yield identical average node degree or
average shortest path length (e.g.~\cite{Costa_surv:2007}).  It is
therefore important to devise different ways of looking at complex
networks.  While random walks have been largely applied in statistical
physics and complex networks, random path-walks (random paths for
short) have received substantially less attention
(e.g.~\cite{Costa_know:2006}).  Paths are important because: (a) they
necessarily terminate (a walk can proceed forever by passing through
repeated edges and nodes); (b) paths are special cases of minimal
spanning trees~\cite{West:2001}, therefore containing the smallest
number of edges required to cover the respective nodes; and (c) they
exist in smaller number than walks, as implied by the requirement of
not repeating edges or nodes.  The current work explored the
possibility to investigate the structural properties of complex
networks as far as two types of paths are concerned: random and
longest.

Random paths are obtained by selecting a node randomly from the
network and initiating a path from that node, until the path
terminates at its both extremities.  The statistical study of the
distribution of type of paths is interesting because the variation of
such structures can reflect intrinsic structural properties in the
networks.  Indeed, the experimental results reported and discussed in
this article indicated that the several types of considered networks
--- namely ER, BA, WS, GG, PA and PN --- are characterized by
distinctive average and standard deviation values of their random path
lengths.  The BA model, in particular, yielded the smallest random
paths, which is explained by the fact that once a hub is visited it
will block access to many other nodes to which it is connected.  The
longest random paths were obtained for the WS model.  Random paths
arising from PN networks tended to present remarkably similar lengths.
The random path lengths tended to increased with the average degree in
all cases.

Largely overlooked in the literature, the longest path of a network
constitutes a particularly interesting global underlying structure.
Though the determination of the longest path in a network represents
an NP-complete, we have shown in the current article --- at least for
relatively small networks ($N=100$) and considered average degrees ---
that a simple algorithm involving the comparison of subsequent random
paths tend to converge to a long path fast and steadily.  Such an
efficiency paves the way to comprehensive investigations of the
distribution and properties of the longest paths in the several
considered network models.  Again, and for similar reasons, the BA
structures tended to present the shortest longest paths, while the WS
networks systematically yielded longest paths comprising all the
nodes.  The longest paths tended to increase with the average degree
in all cases.  At the same time, the longest path lengths were
characterized by surprisingly small standard deviations, except for
the GG model.

Several are the future investigations motivated by the reported
results. First, it would be interesting to investigate the stochastic
distribution of the random and longest paths considering alternative
network models.  It would be also interesting to try to relate
intrinsic random and longest path properties to specific network
measurements in addition to the average degree.  In another direction,
it would be relatively easy to extend the considered concepts and
methods to the analysis of random and longest cycles, which are
closely related to random and longest paths.  The concept of
similarity between paths (as well as walks) can also be refined and
explored further in order to quantify how diverse are the paths
performed in different networks can be.  Such investigations could
provide valuable information about the transient dynamics of the
respective networks.

\begin{acknowledgments}
Luciano da F. Costa thanks CNPq (308231/03-1) and FAPESP (05/00587-5)
for sponsorship.
\end{acknowledgments}

\bibliography{longest_path}
\end{document}